\documentclass{article}

\usepackage{arxiv}

\usepackage[utf8]{inputenc} 
\usepackage[T1]{fontenc}    
\usepackage{hyperref}       
\usepackage{url}            
\usepackage{booktabs}       
\usepackage{amsfonts}       
\usepackage{amsmath}
\usepackage{nicefrac}       
\usepackage{microtype}      
\usepackage{lipsum}
\usepackage{graphicx}

\title{MERCURY: A fast and versatile multi-resolution based global emulator of compound climate hazards}

\author{
 Shruti Nath \\
  Department of Physics\\
  University of Oxford\\
  \texttt{shruti.nath@physics.ox.ac.uk} \\
   \And
 Julie Carreau\\
  Department of Mathematics and Industrial Engineering\\
  Montreal Polytechnique\\
  \texttt{julie.carreau@polymtl.ca} \\
  \And
 Kai Kornhuber \\
  Lamont-Doherty Earth Observatory, Columbia Climate School\\
  and\\
  International Institute for Applied Systems Analysis, Vienna\\
  \texttt{kaik@ldeo.columbia.edu} \\
   \And
 Peter Pfleiderer \\
  Institut für Meteorologie\\
  Universität Leipzig\\
  \texttt{peter.pfleiderer@uni-leipzig.de} \\
   \And
 Carl-Friedrich Schleussner \\
  International Institute for Applied Systems Analysis, Vienna\\
  and\\
  IRI THESys and Geography Faculty, Humboldt-Universität zu Berlin\\
  \texttt{schleussner@iiasa.ac.at} \\
  \And
 Philippe Naveau\\
  Laboratory of Climate and Environmental Sciences\\ Institute Pierre-Simone-Laplace, France\\
  \texttt{philippe.naveau@lsce.ipsl.fr} \\
}

\begin{document}
\maketitle
\begin{abstract}
High-impact climate damages are often driven by compounding climate conditions. For example, elevated heat stress conditions can arise from a combination of high humidity and temperature. To explore future changes in compounding hazards under a range of climate scenarios and with large ensembles, climate emulators can provide light-weight, data-driven complements to Earth System Models. Yet, only a few existing emulators can jointly emulate multiple climate variables. In this study, we present the Multi-resolution EmulatoR for CompoUnd climate Risk analYsis: MERCURY. MERCURY extends multi-resolution analysis to a spatio-temporal framework for versatile emulation of multiple variables. MERCURY leverages data-driven, image compression techniques to generate emulations in a memory-efficient manner. MERCURY consists of a regional component that represents the monthly, regional response of a given variable to yearly Global Mean Temperature (GMT) using a probabilistic regression based additive model, resolving regional cross-correlations. It then adapts a reverse lifting-scheme operator to jointly spatially disaggregate regional, monthly values to grid-cell level. We demonstrate MERCURY’s capabilities on representing the humid-heat metric, Wet Bulb Globe Temperature, as derived from temperature and relative humidity emulations. The emulated WBGT spatial correlations correspond well to those of ESMs and the 95$\%$ and 97.5$\%$ quantiles of WBGT distributions are well captured, with an average of 5$\%$ deviation. MERCURY's setup allows for region-specific emulations from which one can efficiently "zoom" into the grid-cell level across multiple variables by means of the reverse lifting-scheme operator. This circumvents the traditional problem of having to emulate complete, global-fields of climate data and resulting storage requirements.

\end{abstract}


\section{Introduction} \label{sect:intro} 
High impact climatic events are often driven by a combination of physical variables acting together \cite{Raymond2020TheTolerance,Lesk2022CompoundChange}. For example, consequences for human health from extraordinary temperatures are most severe when they coincide with high levels of humidity \cite{Baldwin2023HumiditysDebate}. Multivariate events have been classified as one type of compound events in which correlated or entirely independent variables (or hazards) occur at the same location at the same time, leading to amplification of an impact \cite{Zscheischler2020AEvents}. 
With the aggravated risk of high-impact climate events, there is urgent demand for climate information that allows agile exploration of future climate risks. 

State-of-the-art Earth System Models (ESMs) provide the basis for climate impact studies. However, they are computationally expensive to run and require large storage costs. Moreover, only a small number of their outputs (often post-processed e.g. to specific mean and spread values) are deployed for climate risk assessments, leading to information redundancy. This provides an entry point for low-cost statistical emulators focussing on application relevant output indicators. To date, a wide range of emulators for different applications have been developed, providing yearly to monthly spatially resolved fields of climate variables such as temperature and precipitation, to allow for real-time impact assessments \cite{Alexeeff2018,Beusch2019,Nath2022MESMER-M:Temperature}.  Most emulators however focus on one to two variables at a time \cite{Snyder2019JointDescription,Liu2023GenPhys:Models,Schongart2024IntroducingTemperature,Bassetti2024DiffESM:Models}, or jointly sample multivariate fields based on time sampling approaches. Time sampling approaches however, are hindered under scenarios where not many analogue samples exist leading to repeated sampling of the same fields \cite{Tebaldi2022STITCHES:Simulations}. Moreover, problems of data storage costs and information redundancy, especially when moving to more climate variables, are still present.

In this paper we introduce the Multi-resolutional EmulatoR for CompoUnd climate Risk AnalYsis (MERCURY). MERCURY follows a multi-resolution approach thus representing spatio-temporal climate fields as a series of closed sub-spaces, allowing easy extension to multiple climate variables. This approach condenses key information of large-scale responses under climate change, whilst preserving localized small-scale features. It furthermore allows model reduction when representing the overall mean climate responses across multiplt variables – which has previously been emphasised for climate model emulation \cite{Kitsios2023APathways}. Our approach is built on discrete wavelet-transform methods to compress, estimate and recover our target climate fields of interest \cite{Daubechies1992TenWavelets}. Through a novel lifting scheme, discrete wavelet analysis has furthermore been adapted for irregularly shaped fields \cite{WimSweldens1996THEWAVELETS,Park2022LiftingNetworks} that are more common in climate and geographical spaces (e.g., continents). The lifting scheme is based on a local regression, to split irregularly shaped fields into subspaces and compress them into their key features, whilst maintaining their average values. Recently, \cite{Carreau2023APropagation} demonstrated the use of a lifting scheme based framework to simulate flood wave propagation by proposing an extension to represent spatio-temporal physics-based phenomena. 

The emulator framework proposed in this study is composed of two components. The first component focuses on representing the mean response of impact-relevant regions to GMT. The second component then uses the lifting scheme to "zoom" into and generate the higher resolution, grid-cell level fields. This allows compression of the emulation problem and efficient extension to multivariate representation, conserving both cross-variable and spatial correlations whilst circumventing data storage issues. The structure of this paper is as follows: Section \ref{sect:methods} describes the emulator framework of MERCURY and its evaluation procedure. Following this emulator evaluation results are provided in Section \ref{sect:results}, after which we demonstrate the emulator output in Section \ref{sect:output-example} and proceed to final discussion in Section \ref{sect:disc}.

\begin{figure}[h!] 
\includegraphics[width=14.5cm, height=14.5cm]{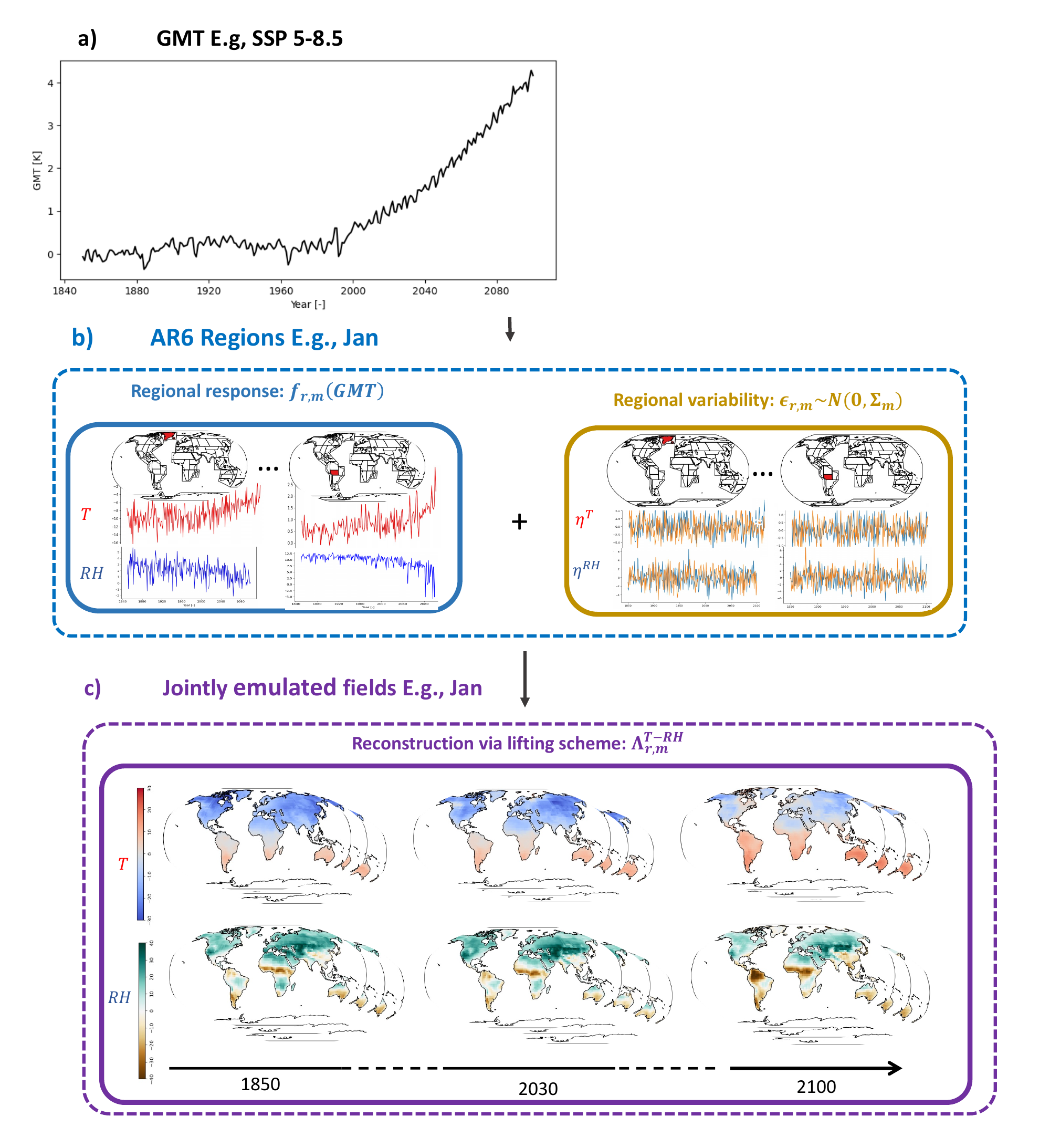}
\caption{MERCURY's framework for generating monthly spatially multivariate climate fields. Yearly GMT values are used as inputs (panel a). The monthly, regional mean response and regional variability for each climate variable is first calculated (panel b). The lifting scheme is then employed to provide monthly spatially resolved, multivariate fields at the grid-cell level (panel c).}
\label{fig:emu_framework}
\end{figure}

\section{Methods}\label{sect:methods}

We are interested in jointly modelling monthly spatially resolved fields for a given set of variables conditional on the yearly Global Mean Temperature ($GMT_y$). MERCURY's framework is summarised in Figure \ref{fig:emu_framework}, and is composed of a regional, component generating monthly ($m$), regional ($r$) mean responses of a climate variable to GMT (described in subsection \ref{sect:det}). Grid-cell level ($gc$) values are then reconstructed from given regional, monthly values using a month-specific lifting scheme (described in subsection \ref{sect:lift}). Data used for training and evaluating MERCURY are described in SI section 1.

\subsection{Representing mean regional monthly responses}\label{sect:det}

To regress the response variable $V$ on the yearly time series of GMT, a classical Additive Model (AM) is applied for each month and for each AR6 region, i.e.
\begin{equation}\label{eq: spline model}
    V_{r,m}= f_{r,m}(GMT_y) + \epsilon_{r,m}, 
\end{equation}
where $m=1, \dots, 12$, $r=1,\dots,44$ and $V_{r,m}$ represents the climatological response during   month $m$ and over region $r$ (either on  temperature or relative humidity). As the relationship between $V_{r,m}$ and $GMT_y$ may not be necessarily linear,  $f_{r,m}(.)$ denotes a smooth   function modelled by a classical cubic spline. The random vector $\epsilon_{r,m}$  corresponds to a zero-mean Gaussian vector with covariance matrix $\Sigma_{r,m}$. This matrix represents monthly regional variability, see  Figures S1 and S2 for visual checks of the Gaussian hypothesis  and serial correlation diagnostics that indicate  only significant positive auto-correlations in tropical latitudes.  
We illustrate \eqref{eq: spline model}, in the left map shown in panel b in Figure \ref{fig:emu_framework} that displays fitted $f_{r,m}$ for $m=1$ (January) and two AR 6 regions with relative humidity in blue  and  temperature in red.
The right map of panel b in Figure \ref{fig:emu_framework} shows the residuals $\epsilon_{r,m}$. 

\subsection{Grid-cell level reconstruction per region}\label{sect:lift}

After fitting model \eqref{eq: spline model} for each month and each AR6 region, the estimated $V_{r,m}$ contains the main response signal and can be used to generate monthly time series of regional values for a given yearly GMT trajectory. To then obtain the corresponding monthly, grid-cell level values for a given region, we employ a lifting scheme based framework, such that:

\begin{equation}\label{eq:reverse:liftingscheme}
    V_{gc,m,y} = \Lambda_{r,m}(V_{r,m})
\end{equation}
Where $gc=1\dots n$ ($n$ representing the number of grid-cells within region $r$),  $\Lambda_{r,m}$ is a month- and region- specific reverse lifting-scheme operator obtained by inverting the lifting scheme. In the following subsections we start by elaborating on the framework of the lifting scheme, followed by how the reverse lifting-scheme operator, $\Lambda_{r,m}$, is obtained from it. Finally, we describe the extension of the reverse lifting-scheme operator, $\Lambda^{T-RH}_{r,m}$, to provide the joint temperature ($T_{gc,m,y}$) and relative humidity ($RH_{gc,m,y}$) emulations.

\subsubsection{Lifting Scheme Framework}

\begin{figure}[h!] 
\includegraphics[width=\textwidth]{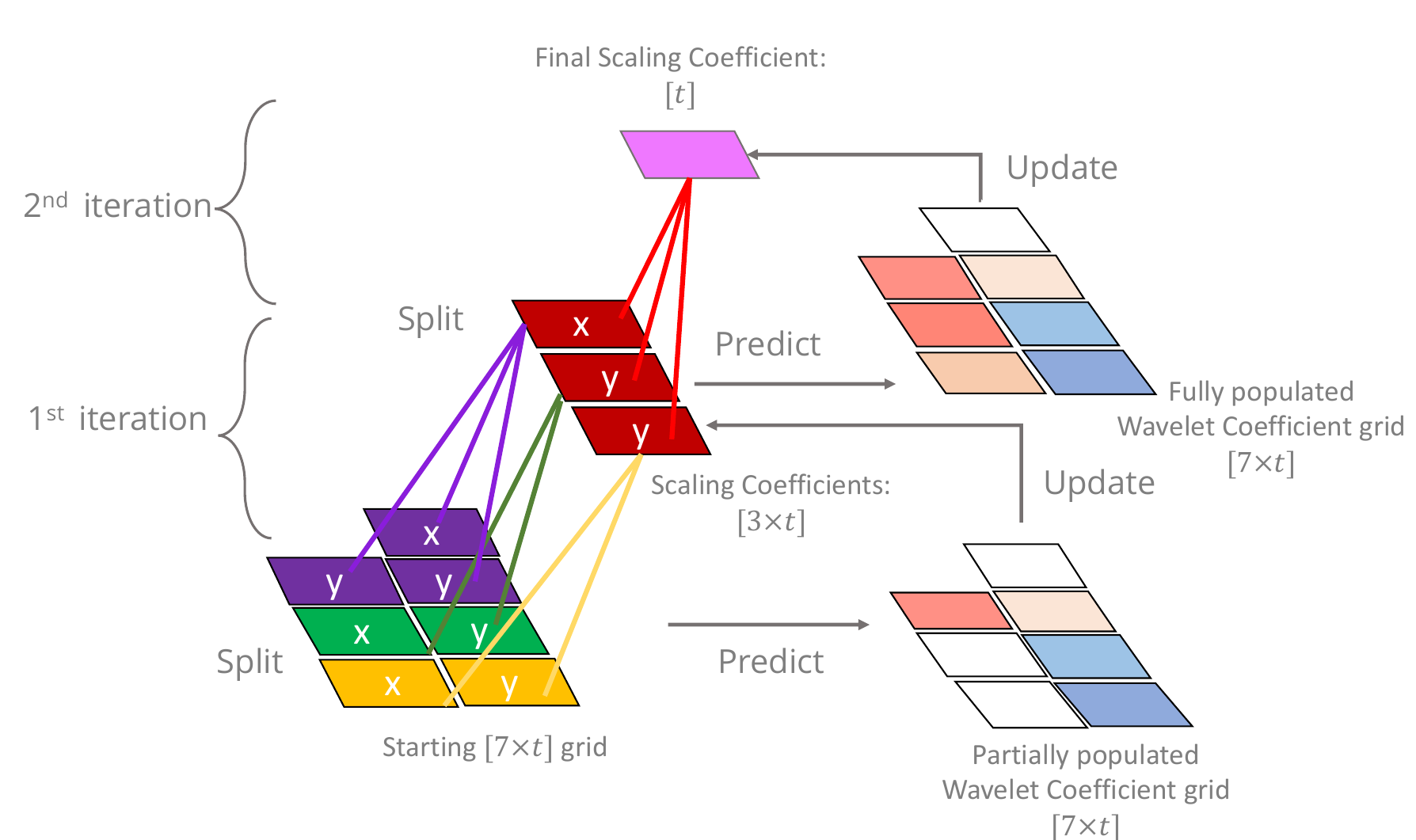}
\caption{Toy example of the lifting scheme applied on a grid consisting of 7 cells with values going from time 1 to $t$. At each iteration of the lifting scheme, the grid is split into groups of two (unless there is an odd number of cells in which case one group of three exists). The 'predict' step stores the wavelet coefficients representing the regression errors resulting from local regression (in our case, naive regression). Finally, the x values are updated with the scaling coefficients obtained by averaging values within each group. Split, predict and update steps are repeated until only a single scaling coefficient exists which corresponds to the grid average, and the wavelet coefficient grid is fully populated up to 6 cells (in the toy example's case two iterations). Within each lifting iteration the grid's spatial dimension is reduced by approximately a half.}
\label{fig:lifting_schematic}
\end{figure}

A month- and region- specific lifting scheme is constructed over the training data by iteratively applying the lifting scheme's split, predict and update steps. These steps had been previously adapted for spatio-temporal datasets by \cite{Carreau2023APropagation}, and are depicted in Figure \ref{fig:lifting_schematic}.

Starting from the original input grid, grid-cells are split into x-y pairs across each latitudinal band, unless an odd number of grid-cells exists in which case one group of three is made. Following this, the wavelet coefficients, $d$ are the errors of a prediction with naive regression as,

\begin{equation}
    d = y-x
\end{equation}
and where a group is of size three, two wavelet coefficients are obtained,

\begin{equation}
    d_1 = y_1-x \hspace{0.3cm}\text{and}\hspace{0.3cm} d_2 = y_2-x .
\end{equation}

Finally, the wavelet coefficients are set aside and the x values in each pair are updated with the scaling coefficients, $c$, corresponding to the average over each group,

\begin{equation}
    c = \frac{x+y}{2} = x + \lambda\cdot d = x + \frac{1}{2}\cdot d \hspace{0.3cm}\text{or}\hspace{0.3cm}c = \frac{x+y_1+y_2}{3} = x + \lambda\cdot (d_1+d_2)= x + \frac{1}{3}\cdot (d_1+d_2),
\end{equation}
where $\lambda$ contains information on the group dimension. 

The split, predict and update steps are iterated through until a single scaling coefficient corresponding to the grid average, and a fully populated grid of wavelet coefficients - except for the cell that corresponds to the scaling coefficient - are obtained. Each iteration corresponds to a resolution level. It should be noted that the $\lambda$ coefficient also needs to be tracked at each iteration, to account for differences in group dimensions being merged during the update step. For example, in the second iteration of the toy example depicted in Figure \ref{fig:lifting_schematic}, the $\lambda$ coefficient would be $\frac{3}{7}$ to account for the differences in the group sizes.

When reversing the lifting scheme, we iterate backwards from the scaling coefficient of the final iteration to the original grid. At each iteration, the x and y values within the corresponding group are obtained as,

\begin{equation}
    x = c-\lambda\cdot d \hspace{0.3cm}\text{and}\hspace{0.3cm} y=d+x = d+ (c-\lambda\cdot d) = c + (1-\lambda)\cdot d .
\end{equation}

As a simple interpretative summary, the lifting framework decomposes a spatio-temporal field into a single time series i.e., scaling coefficient, that contains the spatial average at each time step and its surrounding spatial patterns i.e., wavelet coefficients, that summarise its spatial structures - as deviations from the average - at a given time step. It can thus be thought of as a simpler alternative to an Empirical Orthogonal Function (EOF) analysis based on rudimentary arithmetic that is more flexible in the sense that the decomposition does not make any linearity assumptions (apart from the predict and update operations being locally linear). Alternatively, it could be thought of as a naive Convolutional Network composed of 2-D average pooling layers, but suited for irregular grids with adaptive filter sizes and having deliberate attention focused towards remembering the residual spatial patterns. 

Once, fully iterated over each region and month, the lifting scheme stores, from the last iteration,  a configuration of scaling coefficients (treated as the monthly, regional values), and their corresponding wavelet coefficients (if $n$ is the number of initial grid cells, there are $n-1$ wavelet coefficients). As the reverse lifting scheme starts from the grid average, we may use it on estimated monthly regional values $V_{r,m}$, see \eqref{eq: spline model}, to reconstruct grid-cell level values by selecting suitable wavelet coefficients.

\subsection{Generating new emulations }\label{sect:genemu}

The region- and month- specific reverse lifting-scheme operator $\Lambda_{r,m}$, see \eqref{eq:reverse:liftingscheme}, is able to go back down to a grid-cell level resolution given a single monthly, regional value, interchangeable as the fully lifted scaling coefficient. The key step within the reverse lifting-scheme operator $\Lambda_{r,m}$ is identifying the wavelet coefficients that are prototypical for a given monthly, regional value from which to reconstruct the original climate fields with. To do so, we start by defining the neighbourhood around a given regional, monthly value, $V_{r,m}$ by choosing the twenty scaling values, corresponding to different years, within the lifting scheme's decomposition that are closest to it in value. The associated prototypical wavelet coefficients – referred to as wavelet patterns – are then sampled using Monte Carlo sampling. The Monte Carlo sampling builds a multivariate Gaussian distribution using the 20 wavelet coefficient vectors and samples from that. It assumes normality and stationarity in the wavelet coefficients, and draws from a multivariate Gaussian distribution with covariance matrix constructed across the wavelet coefficients, thus conserving spatial structures within the wavelet coefficients. 

Each sampled wavelet pattern represents a possibility of spatial patterns corresponding to the given regional, monthly value and is used to invert the lifting scheme framework and arrive at a final spatially resolved emulation for that region and month. When doing this using regional monthly values provided across all regions as provided by $f_{r,m}$ (see \eqref{eq: spline model}), we arrive at a single global emulation for a given month. We note that regional cross-correlations are conserved within $f_{r,m}$, and given that the lifting scheme conserves the regional average value, discontinuous jumps between regions are expected to be insignificant. Nevertheless, to ensure seamless blending between regions during reconstruction, we apply a buffer zone around each regional boundary before sampling wavelet coefficients, as previously done to overcome boundary issues for image processing exercises \cite{Hee-Seok2001PolynomialRegression, Naveau2004PolynomialBoundaries}. The buffer zone of width one grid-cell is chosen and the lifting scheme is configured using it. Reverse lifting is conducted using for the whole region plus the buffer zone, after which buffer zone values are simply discarded.

\subsubsection{Reverse lifting-scheme operator for multivariate sampling}

The reverse lifting-scheme operator, $\Lambda_{r,m}$, is extended to a new operator, denoted $\Lambda^{T-RH}_{r,m}$, which enables the joint sampling of $T_{gc,m,y}$ and $RH_{gc,m,y}$. We first identify the key variable that will be used to define the neighbourhood for sampling wavelet coefficients. In our case, we select $T_{r,m,y}$, as its relationship to $GMT_y$ is most established \cite{Tebaldi2014, Tebaldi2018, Herger2015}. Having defined the neighbourhood through the key variable, the usual wavelet sampling steps are then carried out. One difference however, is that the wavelet coefficients can now be jointly sampled across each variable's lifting scheme decomposition, given the mutually defined neighbourhood. To ensure a strict relationship to the key variable, $T_{r,m,y}$, we additionally impose a conditional sampling of the $RH$ wavelet coefficients on the $T$ wavelet coefficients within the Monte Carlo routine, by calculating their conditional covariance matrix,

\begin{equation}
    \Sigma_{RH|T} = \Sigma_{RH,RH}-\Sigma_{RH,T}\cdot\Sigma_{T,T}^{-1}\cdot\Sigma_{T,RH},
\end{equation}
where $\Sigma_{RH|T}$ is the conditional covariance matrix constructed from blocks of the covariance matrix constructed across grid-cells and variables,

\begin{equation}
    \begin{bmatrix}
 \Sigma_{T,T} & \Sigma_{T,RH}\\
\Sigma_{RH,T} & \Sigma_{RH,RH} .
\end{bmatrix}
\end{equation}

A strict conditionality of the distributional space sampled by the Monte Carlo routine on $RH$ to the key variable $T$ is imposed. To this extent, spatially co-occurring and thus compounding patterns are expected to be conserved within the reconstruction process.

\subsection{Evaluation}

In evaluating MERCURY, ten monthly, regional values are produced from $f_{r,m}$, and for each value 100 reconstructions are performed with the reverse lifting-scheme operator $\Lambda^{T-RH}_{r,m}$ resulting in 1000 spatially resolved, monthly, multivariate emulations. MERCURY is evaluated on the test scenario SSP2-4.5. To evaluate the final multivariate $T$ and $RH$ emulations, we first consolidate them into a single representative compound index, the indoor Wet Bulb Globe Temperature ($WBGT$). $WBGT$ is calculated using Stull's method \cite{Stull2011} by first calculating Wet Bulb Temperature ($WBT$):

\begin{equation}
\begin{aligned}
    WBT_{gc,m,y} = T_{gc,m,y}\cdot(c_1\cdot\sqrt{RH_{gc,m,y}+c_2})+ \tan^{-1}(T_{gc,m,y} + RH_{gc,m,y}) - \tan^{-1}(RH_{gc,m,y} - c_3)\\
    + c_4 \cdot RH_{gc,m,y}^{\frac{3}{2}} \cdot\tan^{-1}(c_5 \cdot RH_{gc,m,y})- c_6 ,
\end{aligned}
\end{equation}
where $c_1$, $c_2$, $c_3$, $c_4$, $c_5$ and $c_6$ constants with values 0.16, 8.31, 1.68, 0.0039, 0.023 and 4.69 respectively. $WBGT$ is then obtained from $WBT$ as follows,

\begin{equation}
    WBGT_{gc,m,y} = \frac{2}{3}\cdot WBT_{gc,m,y}+\frac{1}{3}\cdot T_ .
\end{equation}

Final $WBGT$ emulations are inspected for their representation of spatial structures by means of Spearman correlations. For both the ESM and emulator outputs, a Spearman correlation matrix is constructed across all grid cells. The difference is then taken by simply subtracting the emulator's Spearman correlation matrix from the ESM's Spearman correlation matrix. This provides useful diagnosis into how well dominant spatial structures are approximated, where ideally the difference between the two matrices is zero.

We furthermore evaluate MERCURY's representation of the $WBGT$ distributions on a grid-cell and regionally aggregated level. On a grid-cell level, we investigate MERCURY's ability to approximate the median (50$\%$) and the extreme upper (95$\%$ and 97.5$\%$) quantiles of $WBGT$ distributions as calculated from $T$ and $RH$ values outputted by the ESM at each month. We do so by calculating monthly, grid-cell level quantile deviations as used in previous emulator evaluations \cite{Beusch2019, Nath2022MESMER-M:Temperature, Quilcaille2022ShowcasingModels}. Quantile deviations for a quantile, $q_{gc,m}$, are calculated by first calculating quantile time series from the generated emulations for quantile $q$, grid-cell $gc$ and month $m$. The proportion of time steps that the ESM values appear below the emulated quantile value is then calculated ($q_{gc,m}^{ESM}$). The quantile deviation is then $q_{gc,m}^{ESM}-q_{gc,m}$, such that a positive value means that the emulated quantile is larger in value than that of the actual ESM and vice versa. 

We evaluate the  emulator's representation of $WBGT$ distributions on a regionally aggregated level by means of probability rank distributions. Multiple layers of aggregation – calculated using a latitudinally weighted average – from AR6 regions to continental to global are investigated. For each level of aggregation, the probability rank distribution is obtained by calculating the probability rank of the actual, ESM value with respect to the emulated ensemble over all time steps. If the emulated ensemble perfectly captures the actual ESM distribution, we would expect the median probability rank value to correspond to 50$\%$ and so on, such that their final distribution is uniformly distributed.

\section{Results}\label{sect:results}

In the following subsections we first show evaluation results on the test scenario SSP 2-4.5 for the representation of spatial correlations within the multivariate emulations (Section \ref{sect:results-spatver}) and then the representation of the overall distribution (Section \ref{sect:res-regver}).

\subsection{Spatial Evaluation}\label{sect:results-spatver}

\begin{figure}[h!] 
\includegraphics[width=\textwidth]{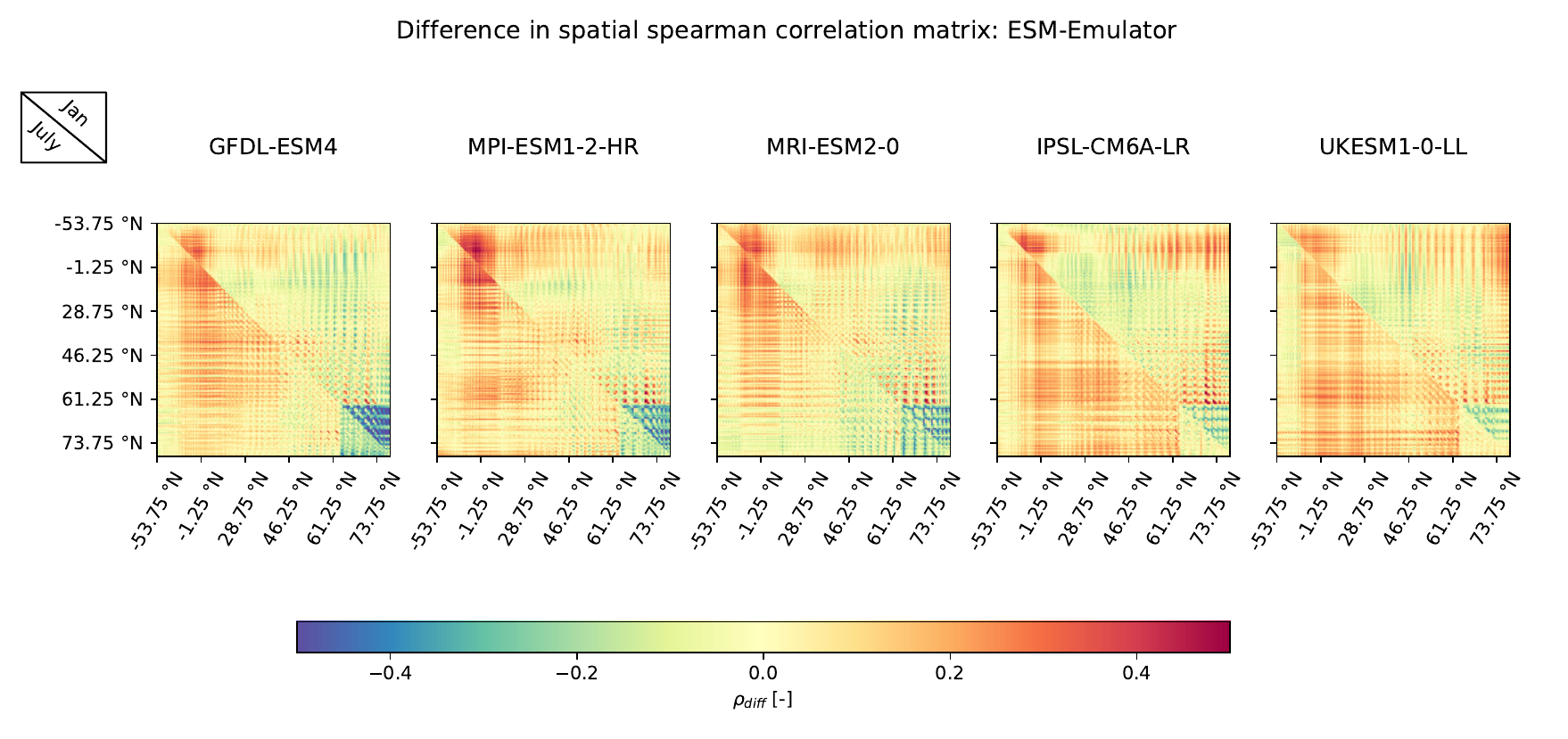}
\caption{Difference between the ESM and emulator Spearman correlation matrix obtained by subtracting the emulator's Spearman correlation matrix from that of the ESM's for test scenario SSP 2-4.5. The Spearman correlation matrix is calculated over all land grid cells for January (upper triangle) and July (lower triangle). }\label{fig:3}
\end{figure}

Differences in the Spearman correlation matrix of the ESM to that of MERCURY are shown in Figure \ref{fig:3}. Across ESMs and months, we note a positive difference, indicating an underestimation of spatial correlations within the emulator. This is particularly apparent in July and concentrated towards correlations with grid cells in equatorial latitudinal bands (i.e. -1.25°N). Since the emulator is a statistical approximation of the ESMs with an approximated spatial covariance matrix, such underestimation of spatial structures is not entirely unexpected. Moreover, equatorial latitudes have stronger temporal correlations (see Durban-Watson test for serial correlation Figure S2), and this indicates some shortcoming in MERCURY's design choice of not accounting for serial correlations. Across all ESMs and months, an overestimation of spatial correlations at latitudes higher that 61.25°N can also be noted. Grid cells above 61.25°N mainly correspond to Greenland, and this could be a product of Greenland being treated as its own AR6 region, leading to overestimation of its regional correlations. 

Comparing the results from Figure \ref{fig:3} to inter-comparison results against the existing monthly temperature emulator MESMER-M (Figure S3), we note that ESMs which had overall poorer CRPSS scores against the benchmark emulator MESMER-M – and therefore no notable improvement in skill – also show larger underestimation of spatial correlations. For further analysis, we focus on 2 ESMs, MRI-ESM2-0 and UKESM1-0-LL, representative of where MERCURY brings meaningful improvements with respect to MESMER-M and where not so much, respectively. We furthermore focus on July as this is the month where the most underestimation of spatial correlations occurs. In case readers are interested in an in-depth analysis of spatial correlations for select grid-cells please refer to SI section 4.

\subsection{Representation of the overall distribution}\label{sect:res-regver}

\begin{figure}[h!] 
\includegraphics[width=\textwidth]{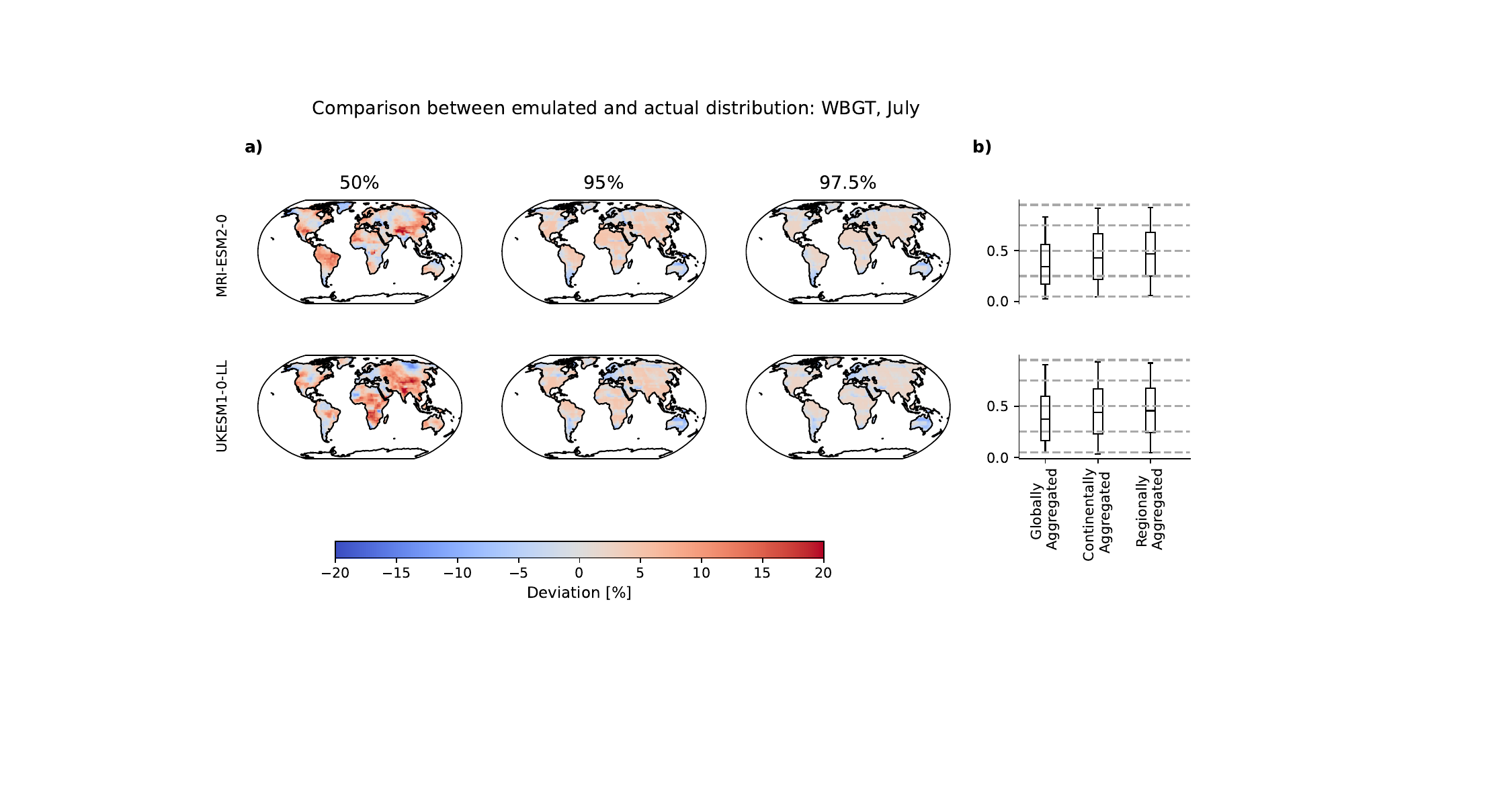}
\caption{Comparison between the emulated and actual distribution for test scenario SSP 2-4.5. a) Quantile deviation maps for the 50$\%$, 95$\%$ and 97.5$\%$ quantiles, where red means that the emulated quantile is warmer than the actual ESM quantile and vice versa for blue. b) Probability rank distributions of the actual data with respect to the emulated ensemble. Data is aggregated to AR6 regional, continental and global levels before calculating the probability ranks. Whiskers indicate the 5th and 95th percentiles. If the distribution of actual data is captured perfectly, then the median should correspond to 0.5, edges to 0.75 and 0.25, and whiskers to 0.05 and 0.95. }\label{fig:4}
\end{figure}

Figure \ref{fig:4} shows 50$\%$, 95$\%$ and 97.5$\%$ quantile deviation maps (panel a) for the two representative ESMs, MRI-ESM2-0 and UKESM1-0-LL, in July. Positive quantile deviations indicate overestimation of the quantile value by MERCURY and vice versa. Whereas  95$\%$ and 97.5$\%$ quantile deviations are quite low (between -5$\%$ and 5$\%$), 50$\%$ quantile deviations show large overestimations with some grid-cells e.g., in South Asia, having values of up to 20$\%$. This indicates that MERCURY may be overestimating the overall mean response of WBGT to GMT on a grid-cell level. 

Probability rank distributions (panel b) look reasonably uniformly distributed for both ESMs across all aggregation levels, albeit globally aggregated showing lower median values (i.e., again an overestimation by MERCURY). In addition to the quantile deviation maps, this indicates that selection of regions could lead to biases in representation of grid-cell level distributions (e.g., perhaps a grid-cell's responses is not so well correlated to the regional response). Consequently, an overall bias in representation of global distributions – as seen in the globally aggregated probability rank distribution – results. Nevertheless, in this study we seek to represent the most impact relevant regions and note that this is a consequence of that design choice.

\section{Example WBGT superensemble time series}\label{sect:output-example}

\begin{figure}[h!] 
\includegraphics[width=\textwidth]{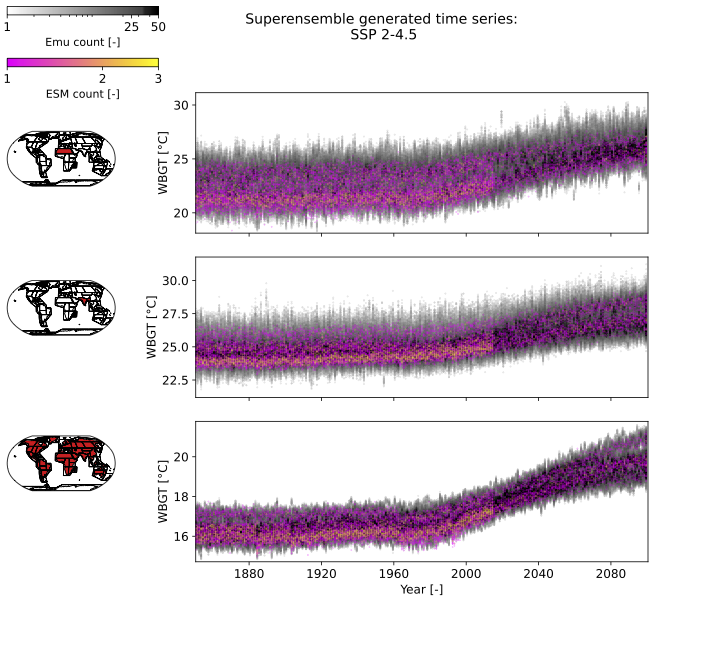}
\caption{2-D histogram time series for the emulated superensemble (greys), with the actual ESM superensemble overlaid for July, SSP 2-4.5. 1000 emulations were produced for each ESM initial-condition ensemble member. Note, that snce historical runs have more ensemble members, the emulation count is also higher up to 2015.}\label{fig:5}
\end{figure}

Figure \ref{fig:5} provides 2-D histogram time series of a superensemble  pooling together 1000 WBGT emulations for each ESM, so a total of 5000 emulations. WBGT values are aggregated to AR6 regions, Sahel and South Asia, and globally. ESM values are also provided for reference.  We again show results for the SSP 2-4.5 scenario. A notable spread and divergence in ESM values across regions and globally is apparent. For example, North Africa displays two modes of ESM WBGT values, starting at approximately 22 \textdegree C and 24 \textdegree C in the year 1850 and increasing at different rates till they converge around 24\textdegree C-25\textdegree C by 2100. MERCURY is able to capture both the spread within each ESM initial-condition ensemble as well as the inter-ESM spread in magnitude and rates of change. This provides useful perspective into the potential of multivariate, lightweight emulators to inform impact assessments. To this extent, they not just provide useful approximations of the ESM initial-condition ensemble spread, but also of the inter-ESM spread, which in some cases may be larger and thus have a greater degree of uncertainty.

\section{Discussion}\label{sect:disc} 
We present MERCURY, a fast and versatile emulator framework that allows approximation of spatially resolved risks from multivariate compound hazards such as wet bulb globe temperature (WBGT). After training on ESM outputs, MERCURY  starts from GMT to deterministically approximate monthly, regional temperature and relative humidity separately by means of a month- and region- specific regression (AM, see \eqref{eq: spline model}). Region-to-region correlations are then added by means of a variable- and month- specific multivariate Gaussian process. Grid-cell level temperatures and relative humidity are jointly reconstructed from their regional values using an operator that reverses the "lifting scheme" adapted for spatio-temporal multivariate sampling. The lifting scheme performs an efficient compression of regional fields based on a local regression,
to iteratively split and compress irregularly shaped fields into their key features. The differences between each iteration are stored as wavelet coefficients, thus enabling reconstruction of the original regional field from a single, regional averaged value. 

The generation of grid-cell level emulations is performed by sampling wavelet coefficients within the "neighbourhood" of a given regional, monthly value from the lifting scheme's decomposition by means of the Monte Carlo method. This allows flexible extension to more variables by sampling their wavelet coefficients from the same neighbourhood. In this study, we select a key variable, temperature, through which to define the neighbourhood which is then used in sampling wavelet coefficients for any additional variables, i.e., relative humidity. In such, we impose a strict, hierarchical dependency of relative humidity on temperature by defining the neighbourhood using temperature only and also by sampling relative humidity using its conditional covariance matrix to temperature. It would be possible to instead have a more egalitarian approach with a mutually defined neighbourhood between variables from which to sample from. In future, a mutually defined neighbourhood could be considered instead which would reduce the computationally complexity of our approach. In line with this, future extension to more climate variables may also encounter non-Gaussian variables such as precipitation. Future extensions of this study may thus benefit from more sophisticated sampling approaches such as use of Diffusion-based Neural Networks, and the framework of MERCURY is modular enough to allow flexible extension for such alternative sampling approaches.

A key advantage of MERCURY lies in its ability to treat irregularly shaped regions. MERCURY doesn't require to derive high resolution information for all regions included, but  one can selectively choose regions of interest to then "zoom" into by means of the reverse lifting-scheme operator. For impact assessments, this means real-time availability of selected impact-relevant regional information without the need to generate and sift through global fields which amount to Peta-bytes of data. To the best of our knowledge, this is the only emulator approach existing so far that tackles the data management problem effectively. MERCURY itself holds low parametric complexity, mainly confined to representing regional responses to GMT, after which it reconstructs grid-cell level responses based on Monte Carlo sampling. This ensures limited growth of parametric uncertainty going down the emulation chain (as otherwise seen in \cite{Nath2024Representing2022}), as well as imposes less stringent functional forms on the grid-cell level responses to GMT.

\clearpage

\section*{Code and Data Availability}
MESMER and MESMER-M are publicly available under Github \cite{Hauser2021MESMER-group/mesmer:0.8.3, Nath2022Snath-xoc/Nath_et_al_ESD_2022_MESMER-M}. Code for the lifting scheme can be found within the GitHub repository: https://github.com/snath-xoc/Lifting. 
The CMIP6 data are available from the public CMIP archive at https://esgf-node.llnl.gov/projects/esgf-llnl/ (ESGF-LNLL, 2022).

\section*{Acknowledgements}
We thank the climate modelling groups listed in Table S1 for producing and making available the CMIP6 model outputs and Urs Beyerle and Lukas Brunner for downloading the CMIP6 data and pre-processing them. SN would like to acknowledge funding from the European Cooperation in Science and Technology, COST action grant CA19139.
JC would like to acknowledge funding by the Natural Sciences and Engineering Research Council of Canada (NSERC) and by Fonds de Recherche du Québec Nature et Technologies (FRQNT).
Part of Naveau’s research work was supported by European H2020 XAIDA (Grant agreement ID: 101003469) and the  French national programs: 80 PRIME CNRS-INSU, Agence Nationale de la Recherche (ANR) under reference ANR-20-CE40-0025-01 (T-REX), the ANR EXSTA, and the PEPR TRACCS programme under grant number ANR-22-EXTR-0005. SN, PP and CFS acknowledge support from the European Union’s Horizon 2020 research and innovation programmes under Grant Agreement No. 101003687 (PROVIDE).

%
%
\bibliographystyle{unsrt}
\bibliography{references}

\begin{thebibliography}{10}

\bibitem{Raymond2020TheTolerance}
Colin Raymond, Tom Matthews, and Radley~M. Horton.
\newblock {The emergence of heat and humidity too severe for human tolerance}.
\newblock {\em Science Advances}, 6(19), 5 2020.

\bibitem{Lesk2022CompoundChange}
Corey Lesk, Weston Anderson, Angela Rigden, Onoriode Coast, Jonas J{\"{a}}germeyr, Sonali McDermid, Kyle~F. Davis, and Megan Konar.
\newblock {Compound heat and moisture extreme impacts on global crop yields under climate change}.
\newblock {\em Nature Reviews Earth {\&} Environment}, 3(12):872--889, 12 2022.

\bibitem{Baldwin2023HumiditysDebate}
Jane~W. Baldwin, Tarik Benmarhnia, Kristie~L. Ebi, Ollie Jay, Nicholas~J. Lutsko, and Jennifer~K. Vanos.
\newblock {Humidity’s Role in Heat-Related Health Outcomes: A Heated Debate}.
\newblock {\em Environmental Health Perspectives}, 131(5), 5 2023.

\bibitem{Zscheischler2020AEvents}
Jakob Zscheischler, Olivia Martius, Seth Westra, Emanuele Bevacqua, Colin Raymond, Radley~M. Horton, Bart van~den Hurk, Amir AghaKouchak, Aglaé J{\'{e}}z{\'{e}}quel, Miguel~D. Mahecha, Douglas Maraun, Alexandre~M. Ramos, Nina~N. Ridder, Wim Thiery, and Edoardo Vignotto.
\newblock {A typology of compound weather and climate events}.
\newblock {\em Nature Reviews Earth {\&} Environment}, 1(7):333--347, 6 2020.

\bibitem{Alexeeff2018}
Stacey~E. Alexeeff, Doug Nychka, Stephan~R. Sain, and Claudia Tebaldi.
\newblock {Emulating mean patterns and variability of temperature across and within scenarios in anthropogenic climate change experiments}.
\newblock {\em Climatic Change}, 146(3-4):319--333, 2018.

\bibitem{Beusch2019}
Lea Beusch, Lukas Gudmundsson, and Sonia~I. Seneviratne.
\newblock {Emulating Earth system model temperatures with MESMER: from global mean temperature trajectories to grid-point-level realizations on land}.
\newblock {\em Earth System Dynamics}, 11(1):139--159, 2 2020.

\bibitem{Nath2022MESMER-M:Temperature}
Shruti Nath, Quentin Lejeune, Lea Beusch, Sonia~I. Seneviratne, and Carl~Friedrich Schleussner.
\newblock {MESMER-M: an Earth system model emulator for spatially resolved monthly temperature}.
\newblock {\em Earth System Dynamics}, 13(2):851--877, 2022.

\bibitem{Snyder2019JointDescription}
Abigail Snyder, Robert Link, Kalyn Dorheim, Ben Kravitz, Ben BondLamberty, and Corinne Hartin.
\newblock {Joint emulation of Earth System Model temperature-precipitation realizations with internal variability and space-time and crossvariable correlation: Fldgen v2.0 software description}.
\newblock {\em PLoS ONE}, 14(10):1--13, 2019.

\bibitem{Liu2023GenPhys:Models}
Ziming Liu, Di~Luo, Yilun Xu, Tommi Jaakkola, and Max Tegmark.
\newblock {GenPhys: From Physical Processes to Generative Models}.
\newblock pages 1--21, 2023.

\bibitem{Schongart2024IntroducingTemperature}
Sarah Sch{\"{o}}ngart, Lukas Gudmundsson, Mathias Hauser, Peter Pfleiderer, Quentin Lejeune, Shruti Nath, Sonia~Isabelle Seneviratne, and Carl-Friedrich Schleu{\ss}ner.
\newblock {Introducing the MESMER-M-TPv0.1.0 module: Spatially Explicit Earth System Model Emulation for Monthly Precipitation and Temperature}, 5 2024.

\bibitem{Bassetti2024DiffESM:Models}
Seth Bassetti, Brian Hutchinson, Claudia Tebaldi, and Ben Kravitz.
\newblock {DiffESM: Conditional Emulation of Temperature and Precipitation in Earth System Models With 3D Diffusion Models}.
\newblock {\em Journal of Advances in Modeling Earth Systems}, 16(10), 10 2024.

\bibitem{Tebaldi2022STITCHES:Simulations}
Claudia Tebaldi, Abigail Snyder, and Kalyn Dorheim.
\newblock {STITCHES: creating new scenarios of climate model output by stitching together pieces of existing simulations}.
\newblock {\em Earth System Dynamics}, 13(4):1557--1609, 11 2022.

\bibitem{Kitsios2023APathways}
Vassili Kitsios, Terence~John O’Kane, and David Newth.
\newblock {A machine learning approach to rapidly project climate responses under a multitude of net-zero emission pathways}.
\newblock {\em Communications Earth {\&} Environment}, 4(1):355, 10 2023.

\bibitem{Daubechies1992TenWavelets}
Ingrid Daubechies.
\newblock {Ten Lectures on Wavelets}.
\newblock PHILADELPHIA, PENNSYLVANIA, 1992. SOCIETY FOR INDUSTRIAL AND APPLIED MATHEMATICS.

\bibitem{WimSweldens1996THEWAVELETS}
{Wim Sweldens}.
\newblock {THE LIFTING SCHEME: A CONSTRUCTION OF SECOND GENERATION WAVELETS}.
\newblock {\em SIAM Journal on Mathematical Analysis}, 1996.

\bibitem{Park2022LiftingNetworks}
Seoncheol Park and Hee-Seok Oh.
\newblock {Lifting Scheme for Streamflow Data in River Networks}.
\newblock {\em Journal of the Royal Statistical Society Series C: Applied Statistics}, 71(2):467--490, 3 2022.

\bibitem{Carreau2023APropagation}
Julie Carreau and Philippe Naveau.
\newblock {A spatially adaptive multi-resolution generative algorithm: Application to simulating flood wave propagation}.
\newblock {\em Weather and Climate Extremes}, 41(May):100580, 2023.

\bibitem{Hee-Seok2001PolynomialRegression}
O.~H. Hee-Seok, Philippe Naveau, and Geunghee Lee.
\newblock {Polynomial boundary treatment for wavelet regression}.
\newblock {\em Biometrika}, 88(1):291--298, 2001.

\bibitem{Naveau2004PolynomialBoundaries}
Philippe Naveau and Hee~Seok Oh.
\newblock {Polynomial wavelet regression for images with irregular boundaries}.
\newblock {\em IEEE Transactions on Image Processing}, 13(6):773--781, 2004.

\bibitem{Tebaldi2014}
Claudia Tebaldi and Julie~M. Arblaster.
\newblock {Pattern scaling: Its strengths and limitations, and an update on the latest model simulations}.
\newblock {\em Climatic Change}, 122(3):459--471, 2014.

\bibitem{Tebaldi2018}
Claudia Tebaldi and Reto Knutti.
\newblock {Evaluating the accuracy of climate change pattern emulation for low warming targets}.
\newblock {\em Environmental Research Letters}, 13(5), 2018.

\bibitem{Herger2015}
Nadja Herger, Benjamin~M. Sanderson, and Reto Knutti.
\newblock {Improved pattern scaling approaches for the use in climate impact studies}.
\newblock {\em Geophysical Research Letters}, 42(9):3486--3494, 2015.

\bibitem{Stull2011}
Roland Stull.
\newblock {Wet-bulb temperature from relative humidity and air temperature}.
\newblock {\em Journal of Applied Meteorology and Climatology}, 50(11):2267--2269, 2011.

\bibitem{Quilcaille2022ShowcasingModels}
Yann Quilcaille, Lukas Gudmundsson, Lea Beusch, Mathias Hauser, and Sonia~I. Seneviratne.
\newblock {Showcasing MESMER‐X: Spatially Resolved Emulation of Annual Maximum Temperatures of Earth System Models}.
\newblock {\em Geophysical Research Letters}, 49(17):1--11, 9 2022.

\bibitem{Nath2024Representing2022}
Shruti Nath, Mathias Hauser, Dominik~L. Schumacher, Quentin Lejeune, Lukas Gudmundsson, Yann Quilcaille, Pierre Candela, Fahad Saeed, Sonia~I. Seneviratne, and Carl-Friedrich Schleussner.
\newblock {Representing natural climate variability in an event attribution context: Indo-Pakistani heatwave of 2022}.
\newblock {\em Weather and Climate Extremes}, 44:100671, 6 2024.

\bibitem{Hauser2021MESMER-group/mesmer:0.8.3}
Mathias Hauser, Lea Beusch, Zebedee Nicholls, and Jonas Schwaab.
\newblock {MESMER-group/mesmer: version 0.8.3}, 2021.

\bibitem{Nath2022Snath-xoc/Nath_et_al_ESD_2022_MESMER-M}
Shruti Nath.
\newblock {snath-xoc/Nath{\_}et{\_}al{\_}ESD{\_}2022{\_}MESMER-M}.
\newblock 4 2022.

\end{thebibliography}


\begin{thebibliography}{10}

\bibitem{Eyring2016}
Veronika Eyring, Sandrine Bony, Gerald~A. Meehl, Catherine~A. Senior, Bjorn Stevens, Ronald~J. Stouffer, and Karl~E. Taylor.
\newblock {Overview of the Coupled Model Intercomparison Project Phase 6 (CMIP6) experimental design and organization}.
\newblock {\em Geoscientific Model Development}, 9(5):1937--1958, 2016.

\bibitem{Stull2011}
Roland Stull.
\newblock {Wet-bulb temperature from relative humidity and air temperature}.
\newblock {\em Journal of Applied Meteorology and Climatology}, 50(11):2267--2269, 2011.

\bibitem{Orlov2019}
Anton Orlov, Jana Sillmann, Asbjørn Aaheim, Kristin Aunan, and Karianne de~Bruin.
\newblock {Economic Losses of Heat-Induced Reductions in Outdoor Worker Productivity: a Case Study of Europe}.
\newblock {\em Economics of Disasters and Climate Change}, 3(3):191--211, 2019.

\bibitem{Baldwin2023HumiditysDebate}
Jane~W. Baldwin, Tarik Benmarhnia, Kristie~L. Ebi, Ollie Jay, Nicholas~J. Lutsko, and Jennifer~K. Vanos.
\newblock {Humidity’s Role in Heat-Related Health Outcomes: A Heated Debate}.
\newblock {\em Environmental Health Perspectives}, 131(5), 5 2023.

\bibitem{ONeill2016}
Brian~C. O'Neill, Claudia Tebaldi, Detlef~P. Van~Vuuren, Veronika Eyring, Pierre Friedlingstein, George Hurtt, Reto Knutti, Elmar Kriegler, Jean~Francois Lamarque, Jason Lowe, Gerald~A. Meehl, Richard Moss, Keywan Riahi, and Benjamin~M. Sanderson.
\newblock {The Scenario Model Intercomparison Project (ScenarioMIP) for CMIP6}.
\newblock {\em Geoscientific Model Development}, 9(9):3461--3482, 2016.

\bibitem{vanVuuren2017EnergyParadigm}
Detlef~P. van Vuuren, Elke Stehfest, David~E.H.J. Gernaat, Jonathan~C. Doelman, Maarten van~den Berg, Mathijs Harmsen, Harmen~Sytze de~Boer, Lex~F. Bouwman, Vassilis Daioglou, Oreane~Y. Edelenbosch, Bastien Girod, Tom Kram, Luis Lassaletta, Paul~L. Lucas, Hans van Meijl, Christoph M{\"{u}}ller, Bas~J. van Ruijven, Sietske van~der Sluis, and Andrzej Tabeau.
\newblock {Energy, land-use and greenhouse gas emissions trajectories under a green growth paradigm}.
\newblock {\em Global Environmental Change}, 42:237--250, 1 2017.

\bibitem{Brunner2020}
Lukas Brunner, Angeline~G. Pendergrass, Flavio Lehner, Anna~L. Merrifield, Ruth Lorenz, and Reto Knutti.
\newblock {Reduced global warming from CMIP6 projections when weighting models by performance and independence}.
\newblock {\em Earth System Dynamics}, 11(4):995--1012, 11 2020.

\bibitem{Iturbide2020AnDatasets}
Maialen Iturbide, José~M. Guti{\'{e}}rrez, Lincoln~M. Alves, Joaquín Bedia, Ruth Cerezo-Mota, Ezequiel Cimadevilla, Antonio~S. Cofi{\~{n}}o, Alejandro Di~Luca, Sergio~Henrique Faria, Irina~V. Gorodetskaya, Mathias Hauser, Sixto Herrera, Kevin Hennessy, Helene~T. Hewitt, Richard~G. Jones, Svitlana Krakovska, Rodrigo Manzanas, Daniel Mart{\'{i}}nez-Castro, Gemma~T. Narisma, Intan~S. Nurhati, Izidine Pinto, Sonia~I. Seneviratne, Bart van~den Hurk, and Carolina~S. Vera.
\newblock {An update of IPCC climate reference regions for subcontinental analysis of climate model data: definition and aggregated datasets}.
\newblock {\em Earth System Science Data}, 12(4):2959--2970, 11 2020.

\bibitem{Fricko2017TheCentury}
Oliver Fricko, Petr Havlik, Joeri Rogelj, Zbigniew Klimont, Mykola Gusti, Nils Johnson, Peter Kolp, Manfred Strubegger, Hugo Valin, Markus Amann, Tatiana Ermolieva, Nicklas Forsell, Mario Herrero, Chris Heyes, Georg Kindermann, Volker Krey, David~L. McCollum, Michael Obersteiner, Shonali Pachauri, Shilpa Rao, Erwin Schmid, Wolfgang Schoepp, and Keywan Riahi.
\newblock {The marker quantification of the Shared Socioeconomic Pathway 2: A middle-of-the-road scenario for the 21st century}.
\newblock {\em Global Environmental Change}, 42:251--267, 1 2017.

\bibitem{Nath2022MESMER-M:Temperature}
Shruti Nath, Quentin Lejeune, Lea Beusch, Sonia~I. Seneviratne, and Carl~Friedrich Schleussner.
\newblock {MESMER-M: an Earth system model emulator for spatially resolved monthly temperature}.
\newblock {\em Earth System Dynamics}, 13(2):851--877, 2022.

\bibitem{Wilks2011StatisticalSciences}
Daniel~S. Wilks.
\newblock {\em {Statistical Methods in the Atmospheric Sciences}}.
\newblock Academic Press, 3rd edition, 2011.

\bibitem{Jolliffe2012ForecastScience}
I.T. Jolliffe.
\newblock {\em {Forecast Verificaton: A Practitioner’s Guide in Atmospheric Science}}.
\newblock Wiley-Blackwell, Oxford, 2nd edition, 2012.

\bibitem{Zamo2018EstimationForecasts}
Michaël Zamo and Philippe Naveau.
\newblock {Estimation of the Continuous Ranked Probability Score with Limited Information and Applications to Ensemble Weather Forecasts}.
\newblock {\em Mathematical Geosciences}, 50(2):209--234, 2 2018.

\end{thebibliography}

\end{document}


\maketitle
\section*{SI 1}
In the analysis, 5 CMIP6 models \cite{Eyring2016} are considered, and we focus on surface air temperature $T$ and relative humidity $RH$, generically referred to as $V$, as these contribute most to hot-humid extremes. For evaluation we use Wet Bulb Globe Temperature calculated from $T$ and $RH$ as a representational compound climate index \cite{Stull2011,Orlov2019,Baldwin2023HumiditysDebate}. Simulations for the SSP5-8.5 high- \cite{ONeill2016} and the SSP1-2.6 low- \cite{vanVuuren2017EnergyParadigm} emission scenarios used as training, thus providing the  full range of emission trajectories between which to interpolate. All available ESM initial-condition ensemble members are used for training. A summary of the CMIP6 models used and their associated modelling groups are given in Table \ref{tab:cmip6}. All ESM runs are obtained at a monthly resolution and are bilinearly interpolated to a spatial resolution of $2.5°\times2.5°$ \cite{Brunner2020}. 

MERCURY is trained using yearly GMT as input, which is calculated using a latitudinally weighted average. All climate variables are taken as anomalies with respect to the annual climatological mean over the reference period of 1850–1900. As further described in the main test methods section, MERCURY consists of a regional component for which we use the AR6 regions \cite{Iturbide2020AnDatasets}. The AR6 regions are the most impact relevant, having been identified by the Intergovernmental Panel for Climate Change (IPCC) as the standard reference regions for subcontinental analysis. The final evaluation of multivariate emulations was conducted on the test SSP2-4.5 medium-emission scenario \cite{Fricko2017TheCentury}.
\label{app:a}
\begin{table}
\caption{List of the 5 employed CMIP6 models, the modelling groups providing them}
\label{tab:cmip6}
 \begin{tabular}{cc}
Model & Modelling centre (or group)\\
\hline
GFDL-ESM4 & NOAA Geophysical Fluid Dynamics Laboratory\\
IPSL-CM6A-LR & Institut Pierre-Simon Laplace\\
MPI-ESM1-2-HR & Max-Planck-Institut für Meteorologie (Max Planck Institute for Meteorology)\\
MRI-ESM2-0 & Meteorological Research Institute NESM3\\
UKESM1-0-LL & Met Office Hadley Centre\\
\hline
\end{tabular}
\end{table}

\section*{SI 2}
\begin{figure}[h!] 
\includegraphics[width=\textwidth]{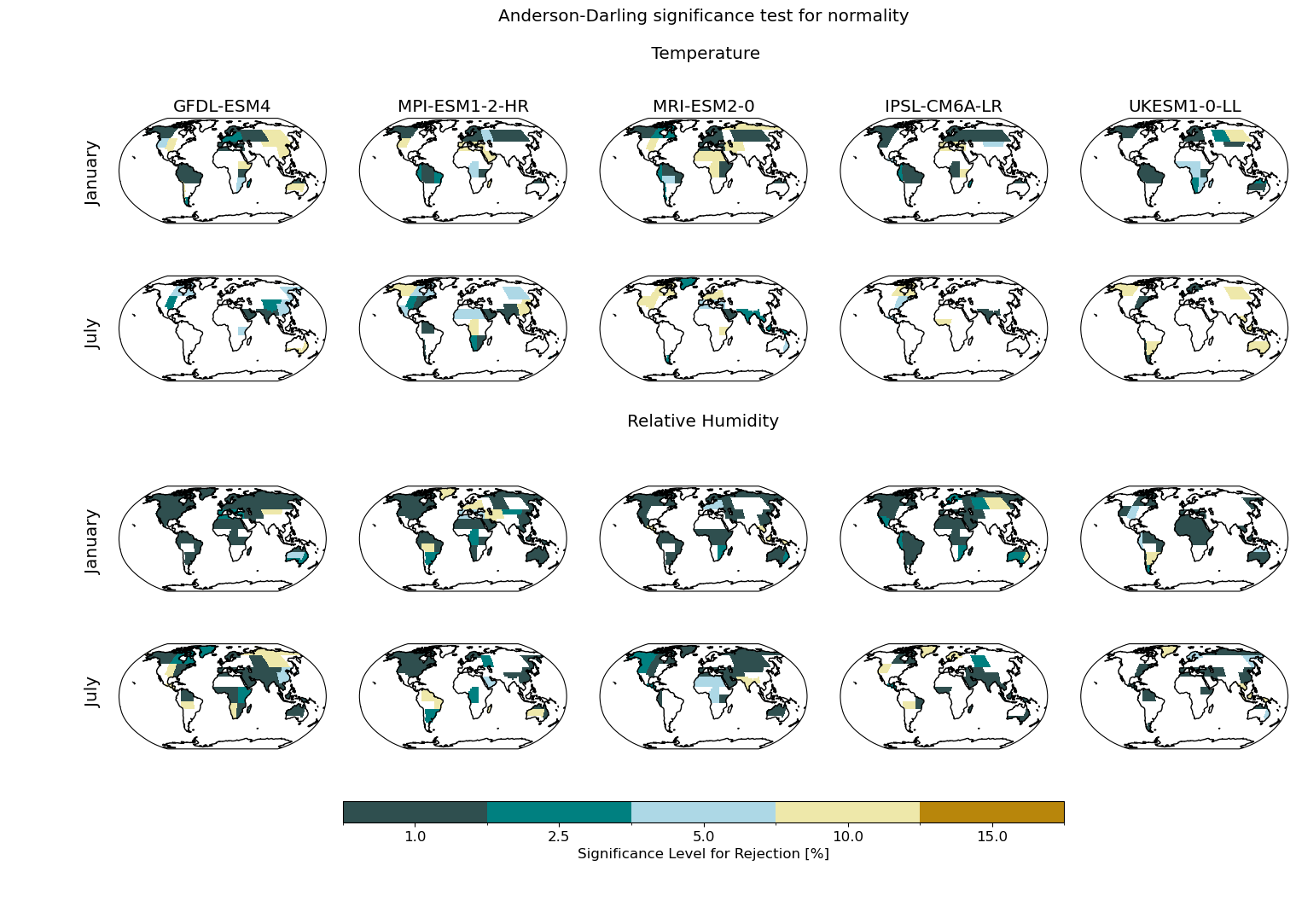}
\caption{Anderson-Darling test for normality performed on the monthly, regional residuals from the AM (see Section 2.1) for Temperature (top panel) and Relative Humidity (bottom panel) and months January (first row) and July (second row). Colours indicate the significance level at which the assumption of normality can be rejected, generally a 2.5$\%$ signficance level or below is considered signficant in this study.}\label{fig:B1}
\end{figure}

\begin{figure}[h!] 
\includegraphics[width=\textwidth]{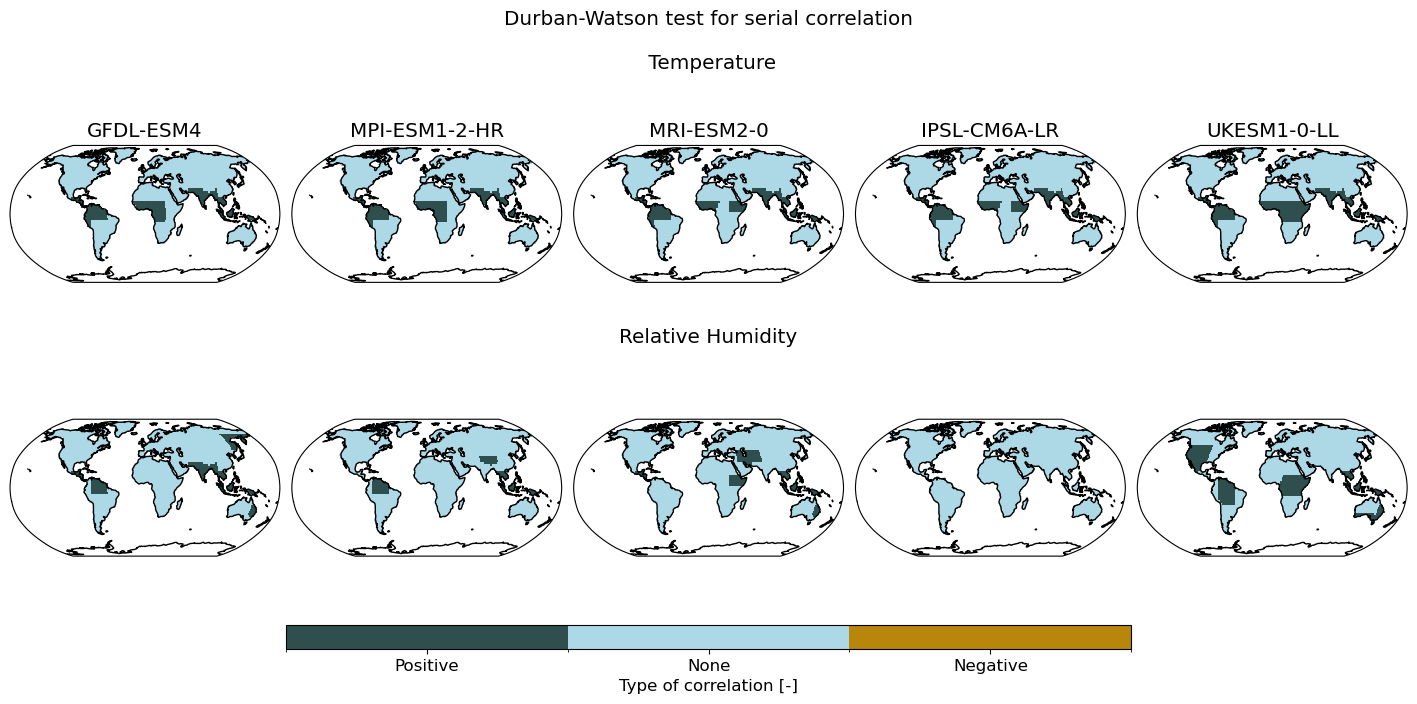}
\caption{Durban-Watson test for stationarity in the monthly, regional residuals from the GAM (see Section 2.1) for Temperature (top panel) and Relative Humidity (bottom panel). Positive (negative) means a significant positive (negative) trend detected in the residuals.}\label{fig:B2}
\end{figure}

\section*{SI 3} 
MERCURY is benchmarked 
against an existing monthly emulator, MESMER-M \cite{Nath2022MESMER-M:Temperature}. Readily available MESMER-M temperature emulations for SSP5-8.5 were used for this. A full description of MESMER-M can be found under \cite{Nath2022MESMER-M:Temperature}, and some key differences between MESMER-M and this study's approach are noted in Table \ref{tab:difftomesmer}.
\begin{table}
\caption{Benchmarking against MESMER-M: SSP5-8.5, surface air temperature}
\label{sect:methods-bench}
\label{tab:difftomesmer}
 \begin{tabular}{c|p{0.35\linewidth}|p{0.35\linewidth}}
 & MESMER-M & MERCURY\\
\hline
Input Variable & yearly, local temperature & yearly, Global Mean Temperature (GMT)\\
\hline
Emulated variables & local, monthly temperature & local, monthly temperature and relative humidity\\
\hline
Deterministic component & local harmonic model & regional Generalised Additive Model (GAM)\\
\hline
Temporal variability & accounted for via local, month-specific AR(1) process & unaccounted for\\
\hline
Small-scale spatial patterns & accounted for via multivariate Gaussian process with localised spatial covariance matrix & accounted for within sampling routine of lifting scheme\\
\hline
Large-scale spatial patterns & loosely accounted for (depending on degree of localisation employed within spatial covariance matrix) & accounted for within regional variability component\\
\hline
\end{tabular}
\end{table}
Benchmarking is carried out using MESMER-M emulations as a reference ($ref$) and the historical and SSP 5-8.5 scenarios. The Continous Rank Probability Skill Score (CRPSS) is used,

\begin{equation}
    CRPSS = 1-\frac{CRPS}{CRPS_{ref}}
\end{equation}

Where CRPS refers to the Continuous Rank Probability  \cite{Wilks2011StatisticalSciences, Jolliffe2012ForecastScience, Zamo2018EstimationForecasts} and is calculated for each month as,

\begin{equation}
\begin{aligned}
    CRPS(V_{m,y},V^{ESM}_{m,y})_m = 
 \frac{1}{N_y}\Sigma^{y=N_y}_{y=1}[\int^{+\infty}_{-\infty} \frac{1}{N_{e}}
    \Sigma^{e=N_e}_{e=1}[\mathbb{1}(V\geq V_{m,y,e})-
    \mathbb{1}(V\geq V^{ESM}_{m,y})]^2 dV]
\end{aligned}
\end{equation}

\begin{figure}[h!] 
\includegraphics[width=15cm]{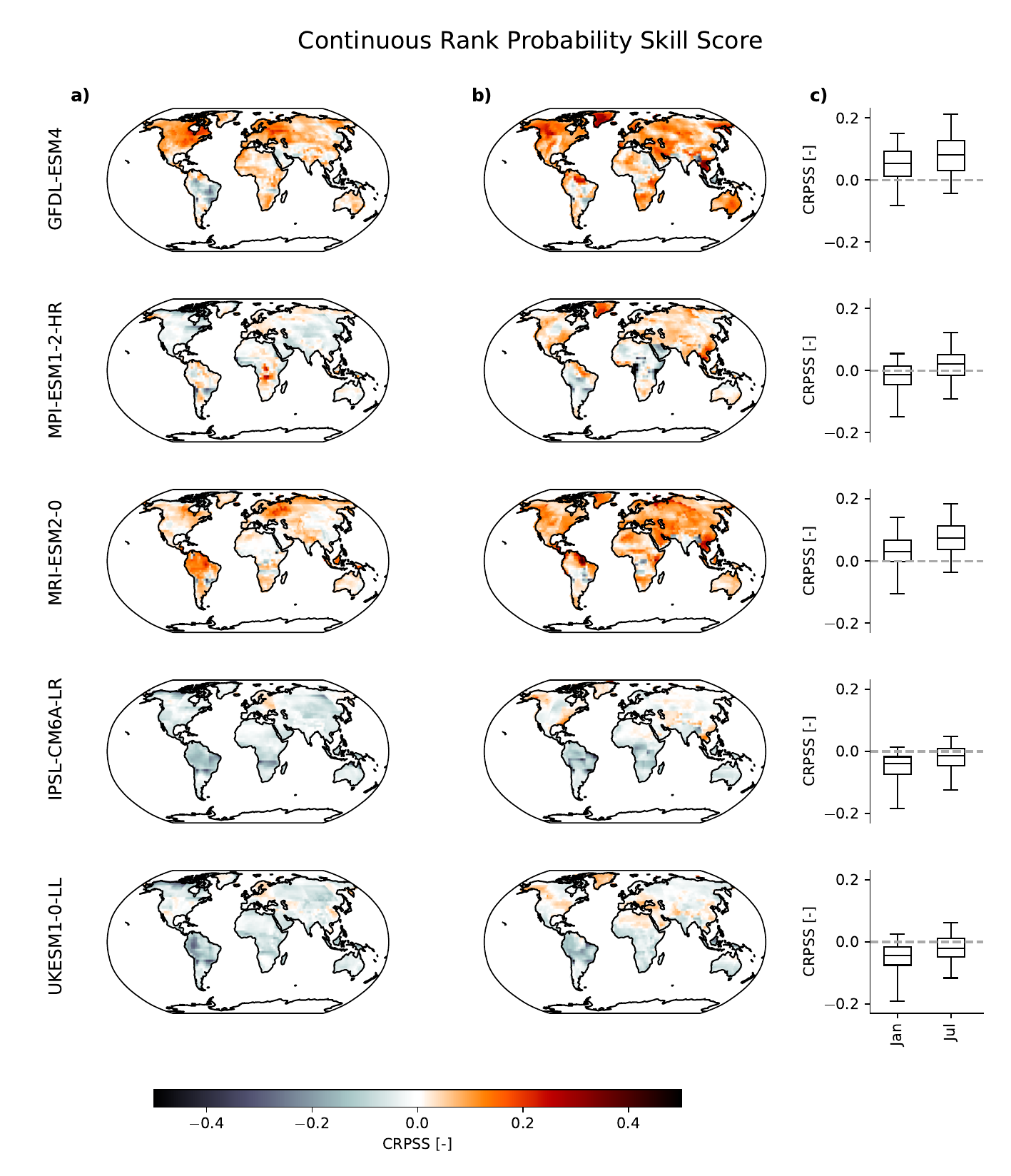}
\caption{Continuous Rank Probability Skill Score for all ESMs (rows) using MESMER-M emulations as a reference. a) and b) provide maps of CRPSS for January and July respectively, where a value above zero (red) indicates better performance as compared to the reference and vice versa for below zero (black), c) provides box plots summarising the overall global spread in CRPSS values for January and July, whiskers indicate the 5$\%$ and 95$\%$ quantiles.}\label{fig:crpss_1}
\end{figure}

If the proposed emulator performances are better than the reference, CRPSS values will be positive and vice versa. CRPSS scores are shown in Figure \ref{fig:crpss_1}. January and July show an overall improvement brought by this study's emulator for GFDL-ESM4, MPI-ESM1-2-HR and MRI-ESM2-0, with CRPSS boxplots (column c) showing at least 75$\%$ of grid points with positive CRPSS scores (except for MPI-ESM2-HR in January). In contrast, IPSL-CM6A-LR and UKESM1-0-LL show minimal improvement with at least 75$\%$ of grid points having zero to negative CRPSS scores in both January and July. Nevertheless, median CRPSS scores are close to 0 indicating a median global performance similar to MESMER-M. Given that this study's emulator only takes GMT as information, whereas MESMER-M takes local, yearly temperatures information, the performance gain weighed against the reduced model complexity is still noteworthy.

\section*{SI 4}
\begin{figure}[h!] 
\includegraphics[width=\textwidth]{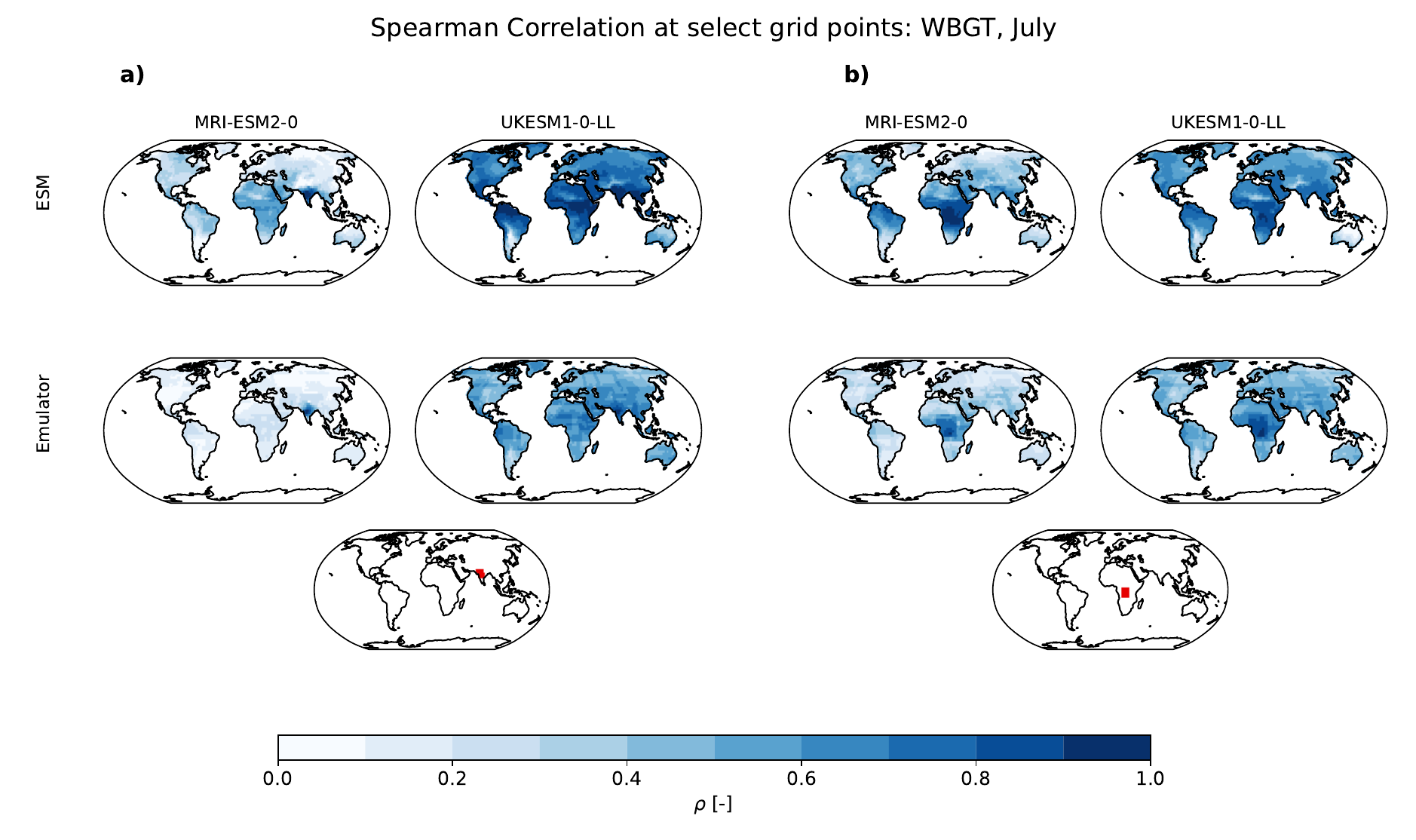}
\caption{Spearman correlations calculated for two select grid points (in red) shown in panels a) and b), for test scenario, SSP 2-4.5. Upper and lower rows in each panel show correlations for ESM and the emulator data respectively. Emulator correlations are calculated for each emulation individually and then averaged, 1000 emulations were used.}\label{fig:D1}
\end{figure}

Figure \ref{fig:D1} contrasts ESM and emulator maps of spatial correlations for the two representative ESMs and select grid points in South Asia (column a) and Central Africa (column b). In both ESMs, it is apparent that while MERCURY represents regional correlations well, it underestimates large-scale correlations, more so for UKESM1-0-LL. This indicates that the region-to-region correlations are underestimated within the AM, whereas the reverse lifting-scheme operator is able to conserve the local correlations within a region. It should be noted however that UKESM1-0-LL displays much stronger and larger spatial correlations as compared to MRI-ESM2-0. The difference between ESM and emulator spatial correlations may thus be minimal as compared to the difference between ESM to ESM spatial correlations. 

\bibliographystyle{unsrt}
\bibliography{references}